\documentclass[aip,apl,amsfonts,amssymb,amsmath,groupedaddress,twocolumn]{revtex4}

\usepackage{graphicx}
\usepackage{amsmath}
\usepackage{amsfonts}
\usepackage{amssymb}
\usepackage{graphicx}
\usepackage{epstopdf}

\begin{document}
\title{Modeling ballistic effects in frequency-dependent transient thermal transport using diffusion equations}

\author{Jesse Maassen}
\email{jmaassen@dal.ca}
\affiliation{Department of Physics and Atmospheric Science, Dalhousie University, Halifax, Nova Scotia, Canada, B3H 4R2}
\author{Mark Lundstrom}
\affiliation{School of Electrical and Computer Engineering, Purdue University, West Lafayette, IN 47907, USA}

\begin{abstract}
Understanding ballistic phonon transport effects in transient thermoreflectance experiments and explaining the observed deviations from classical theory remains a challenge. Diffusion equations are simple and computationally efficient but are widely believed to break down when the characteristic length scale is similar or less than the phonon mean-free-path. Building on our prior work, we demonstrate how well-known diffusion equations, namely the hyperbolic heat equation and the Cattaneo equation, can be used to model ballistic phonon effects in frequency-dependent periodic steady-state thermal transport. Our analytical solutions are found to compare excellently to rigorous numerical results of the phonon Boltzmann transport equation. The correct physical boundary conditions can be different from those traditionally used and are paramount for accurately capturing ballistic effects. To illustrate the technique, we consider a simple model problem using two different, commonly-used heating conditions. We demonstrate how this framework can easily handle detailed material properties, by considering the case of bulk silicon using a full phonon dispersion and mean-free-path distribution. This physically transparent approach provides clear insights into the nonequilibrium physics of quasi-ballistic phonon transport and its impact on thermal transport properties.    
\end{abstract}


\maketitle

\section{Introduction}
Recent transient experiments probing the thermal transport properties of materials on short length- and/or time-scales have reported deviations from expected classical theory, which often corresponds to a reduction in extracted thermal conductivity \cite{Koh2007,Siemens2010,Minnich2011a,Johnson2013,Regner2013,Wilson2014}. This has largely been attributed to quasi-ballistic phonon transport, arising when the phonon mean-free-path (MFP) is similar to or greater than the characteristic length scale in the experiment. Traditional diffusion heat equations are commonly relied upon for analyzing raw data and extracting thermal properties, but the classical heat equations are widely believed to break down under conditions of non-diffusive transport. This paper addresses the need for fast and accurate techniques to analyze transient thermal measurements.

To capture ballistic behavior, theoretical efforts have largely focused on using rigorous approaches such as the phonon Boltzmann transport equation (BTE). Such detailed numerical studies \cite{Minnich2011b,Peraud2011,Peraud2012,Collins2013,Ding2014,Zeng2014} can be too computationally demanding for the routine analysis of experiments. The phonon BTE can, however, also provide a starting point for deriving simple models to elucidate the relationship between the intrinsic phonon properties of a material (e.g. phonon MFP) and the measured thermal transport characteristics \cite{Maznev2011,Wilson2013,Regner2014,Vermeersch2014a,Vermeersch2014b,Hua2014a,Hua2014b,Yang2015,Hua2015}. These simple models are often problem-specific -- assuming simplified geometries and material structures. Moreover they are not straightfowardly compatible with traditional analysis approaches, and must often be used as post-processing tools to analyze the extracted thermal properties. Ideally one would like to have an approach that captures ballistic effects, but that can also be applied to a wide class of problems, readily handle material structures similar to the experimental setup and that can be used to analyze raw data. Such a technique is described in this paper.

In our previous work, we showed that steady-state and transient diffusion equations can capture ballistic phonon effects as long as the correct physical boundary conditions are used. When properly implemented, diffusion equations provide good agreement with the phonon BTE \cite{Maassen2015a,Maassen2015b}. In this paper, we extend previous work to the periodic steady-state case and analyze ballistic effects in model transient thermoreflectance experiments using diffusion equations. We compare diffusion equation solutions to recently-reported rigorous numerical results of the phonon BTE and find excellent agreement. We also demonstrate how this approach readily supports the inclusion of detailed phonon properties, including a full phonon dispersion and MFP distribution. Finally, as an illustration of the technique, we consider a simple model problem using two different, commonly-used heating conditions. Not surprisingly, we find that the different cases produce different results, but we also show that the quantities that would be measured in an experiment are insensitive to the specific heating condition, at least for this simple, model problem. The main conclusion of this work is that the range of problems that can be addressed with diffusion equations is much broader than has been generally understood.  

The paper is outlined as follows. Section \ref{sec:theory} presents the problem under consideration and describes our theoretical approach. Section \ref{sec:results} shows our solutions for the model structure, compares our solutions to those obtained from the phonon BTE, and applies the technique to bulk silicon using detailed material properties. Section \ref{sec:discussion} discusses our results and the relation to experiments. Finally, in Section \ref{sec:summary} we summarize our findings.

\section{Model Structure and Theoretical Approach} 
\label{sec:theory}
In this work we model a simple structure comprised of a semi-infinite ($0$$<$$x$$<$$\infty$) semiconductor/insulator slab, driven by periodic harmonic heating at the surface ($x=0$). Actual structures in thermoreflectance experiments are more complicated \cite{Koh2007,Regner2013,Wilson2014}, and modeling such experiments requires considering, for example, the metal transducer and the finite size of the heating source in the $y$-$z$ plane at the surface. We chose a simple structure to more easily illustrate and analyze the role of ballistic phonon effects, to demonstrate how such effects are captured by diffusion equations, and to compare to rigorous numerical solutions of the phonon BTE \cite{Regner2013,Yang2015}. 

With time-domain thermoreflectance (TDTR) the heating at the surface is driven by a train of short laser pulses ($\sim$ ps or fs) that are modulated at a given frequency ($f\sim$ 1-10 MHz), while with frequency-domain thermoreflectance (FDTR) the heating at the surface is generated by a modulated continous laser stream ($f\sim$ 1-100 MHz). Here, we consider the simpler case of periodic harmonic heating, similar to the conditions of FDTR, although it is possible to also calculate the TDTR response since both are mathematically connected \cite{Collins2014}.

We begin with the McKelvey-Shockley flux method \cite{Mckelvey1961,Shockley1962}, which was shown to treat phonon transport from the ballistic to diffusive transport regime \cite{Maassen2015b}: 
\begin{align}
\frac{1}{v_x^+} \frac{\partial I_Q^+}{\partial t} + \frac{\partial I_Q^+}{\partial x} &= -\frac{I_Q^+}{\lambda}+\frac{I_Q^-}{\lambda}, \label{mk_flux1} \\
-\frac{1}{v_x^+} \frac{\partial I_Q^-}{\partial t} + \frac{\partial I_Q^-}{\partial x} &= -\frac{I_Q^+}{\lambda}+\frac{I_Q^-}{\lambda}, \label{mk_flux2}
\end{align}
where $I_Q^{\pm}(x,t,\epsilon)$ are the forward/backward heat fluxes, $\lambda(\epsilon)$ is the mean-free-path for backscattering, $v_x^+(\epsilon)$ is the average $x$-projected velocity, and $\epsilon$ is the phonon energy. The net heat current and heat density are given by $I_Q=I_Q^+ - I_Q^-$ and $Q = ( I_Q^+ + I_Q^- )/ v_x^+$. Eqns. (\ref{mk_flux1})-(\ref{mk_flux2}) have been derived assuming each phonon energy, or equivalently phonon frequency ($\nu = \epsilon/h$), is independent (i.e. scattering treated at the level of relaxation time approximation), and that the angle-dependent $x$-projected phonon velocity distribution is approximated by the angle-averaged value, $v_x^+$. Although our approach is written in terms of phonon energy $\epsilon$, if preferred, it is possible and equivalent to deal with phonon frequency by making the substitution $\epsilon\rightarrow h \nu$. 

Under conditions of small temperature variations, the McKelvey-Shockley equations can be rewritten exactly as the hyperbolic heat equation (HHE) and the Cattaneo equation \cite{Maassen2015b}:
\begin{align}
\frac{\partial T}{\partial t} + \tau_Q \frac{\partial^2 T}{\partial t^2} &= \frac{\kappa}{C_V}\frac{\partial^2 T}{\partial x^2}, \label{hhe} \\
I_Q + \tau_Q \frac{\partial I_Q}{\partial t}   &=  -\kappa \frac{\partial T}{\partial x}, \label{cattaneo}
\end{align}
where $T$ is the temperature, $\tau_Q = \lambda/(2v_x^+)$ is the heat relaxation time, $\kappa = C_V \lambda v_x^+/2$ is the bulk thermal conductivity and $C_V$ is the heat capacity. The equivalence between this expression for bulk thermal conductivity and the classic relation is shown in Appendix B of Ref. \cite{Maassen2015a}. (Note that the temperature in these equations is the average of the temperature of the forward and reverse heat fluxes.) These equations are modified versions of the heat equation and Fourier's law that capture finite-velocity propagation \cite{Cattaneo1958}. While these diffusion equations are widely believed to break down when ballistic effects are present, we showed that this is not the case \cite{Maassen2015a,Maassen2015b}. Here we will solve the HHE and Cattaneo eq. to study the role of ballistic effects in frequency-dependent thermal transport.

Since periodic harmonic oscillations are driving the heating at the surface, one can show that the solutions for temperature and heat current will have the form $T(x,t)=T(x)e^{i\omega t}$ and $I_Q(x,t)=I_Q(x)e^{i\omega t}$, respectively. Inserting these expressions in the HHE and Cattaneo eq., we obtain the following new equations:
\begin{align}
\frac{\kappa}{C_V}\frac{\partial^2 T}{\partial x^2} &- i\omega \left(1+i\omega \tau_Q\right) T = 0, \label{hhe_2} \\
I_Q &=  -\frac{\kappa}{\left(1+i\omega \tau_Q\right)} \frac{\partial T}{\partial x}, \label{cattaneo_2}
\end{align}
which can be viewed as modified versions of the heat equation and Fourier's law, and as $\tau_Q$$\rightarrow$$0$ we retrieve these classic expressions. We can see how the heat current will be reduced by the denominator at high frequency compared to that predicted by Fourier's law. We will show that by solving these simple diffusion equations, we can obtain excellent agreement compared to numerical results of the phonon BTE without parameter adjustment, as shown in Fig. \ref{fig2}(b)-(e) and Fig. \ref{fig3} (solid lines: our approach, markers: phonon BTE).

\subsection{Boundary conditions}
\label{sec:BCs}
Here we present the correct physical boundary conditions, equivalent to those from the phonon BTE, to be used with our approach based on diffusion equations (i.e. the HHE and Cattaneo eq.). Two types of periodic heating at the surface have been used to study ballistic phonon effects. Case I: the surface is in contact with an ideal reservoir with an oscillating temperature $T(t)=\Delta T e^{i\omega t}$, where $\omega$ is the angular frequency \cite{Regner2014,Yang2015}. Case II: there is an oscillating heat current at the surface $I_Q(t)=I_Q^0  e^{i\omega t}$ \cite{Regner2013}. In both cases, another boundary condition imposes that the excess temperature variations decay to zero as $x$$\rightarrow$$\infty$.

We previously showed that implementing the correct physical boundary conditions, that is the boundary conditions imposed on the directed heat fluxes $I_Q^{\pm}$, is key to capturing ballistic effects. We find the correct physical boundary conditions at $x=0$ are
\begin{align}
T(0^+,t) - \frac{\lambda}{2\left(1+i\omega\tau_Q\right)}\left.\frac{\partial T}{\partial x}\right|_{0^+} &= \Delta T e^{i\omega t}, \label{bc:temp} \\
I_Q(0^+,t) &= I_Q^0 e^{i \omega t}, \label{bc:iq}
\end{align}
for the case of $T$-controlled (case I) and $I_Q$-controlled (case II) heating at the surface, respectively. Appendix \ref{app:BC} shows how these boundary conditions are obtained. Note that a traditional approach would not have the second term on the left-hand side of Eq. (\ref{bc:temp}), which is responsible for capturing temperature jumps at the surface. In what follows, ``traditional approach" refers to solving the heat equation and Fourier's law using the classical boundary conditions, i.e. $T(0^+,t)=\Delta T e^{i\omega t}$ or $I_Q(0^+,t)= I_Q^0 e^{i\omega t}$.

\section{Results} 
\label{sec:results}

\subsection{Analytical solutions}
\label{sec:anal_soln}
Solving the HHE and Cattaneo eq., Eqns. (\ref{hhe_2})-(\ref{cattaneo_2}) with the appropriate boundary conditions, we obtain analytical solutions for the temperature and heat current distributions. For the case of a temperature-controlled surface ($T(t)=\Delta T e^{i \omega t}$, given by Eq. (\ref{bc:temp})), we find 
\begin{align}
 T(x,t)&=\Delta T \left[ \frac{1+i\omega \tau_Q}{(1+i\omega \tau_Q)+\lambda k /2}  \right] e^{-k x} e^{i\omega t}, \label{sol:temp_bc1} \\
 I_Q(x,t)&= \Delta T \left[ \frac{\kappa k}{(1+i\omega \tau_Q)+\lambda k /2}  \right] e^{-k x} e^{i\omega t}, \label{sol:iq_bc1}
\end{align}
where $k(\omega)$ is expressed as
\begin{align}
 k(\omega)&=\sqrt{\frac{2i\omega}{\lambda v_x^+} \left( 1+i\omega \tau_Q  \right)}. \label{sol:k} 
\end{align}
For the case of a heat current-controlled surface ($I_Q(t)=I_Q^0 e^{i\omega t}$, given by Eq. (\ref{bc:iq})), we find
\begin{align}
 T(x,t)&=I_Q^0 \left[ \frac{1+i\omega \tau_Q}{\kappa k}  \right] e^{-k x} e^{i\omega t}, \label{sol:temp_bc2} \\
 I_Q(x,t)&=I_Q^0  e^{-k x} e^{i\omega t}. \label{sol:iq_bc2} 
\end{align}

\begin{figure}	
\includegraphics[width=8.5cm]{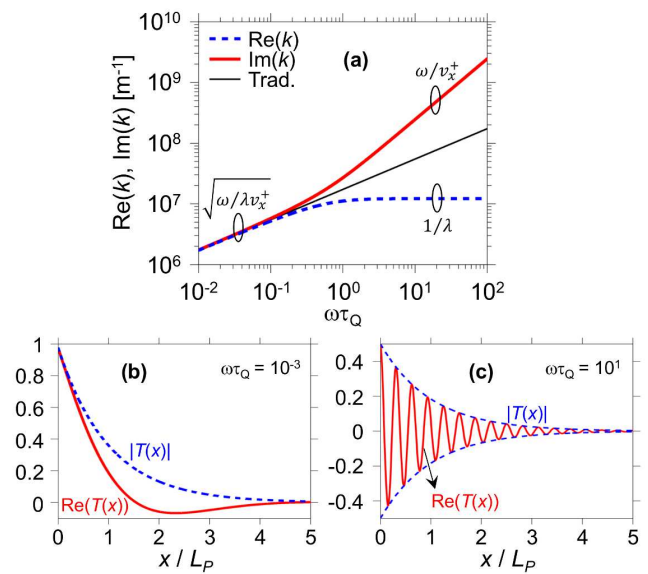}
\caption{(a) Real and imaginary parts of $k$ wavevector versus frequency $\omega \tau_Q$. Traditional approach corresponds to solving the heat equation. (b)-(c) Magnitude and real parts of the normalized temperature profile $T(x)$ versus normalized position $x/L_P$ for $\omega \tau_Q$=10$^{-3}$ (b) and 10$^1$ (c), where $L_P$=Re($k$) is the penetration depth. The adopted parameters are taken from Ref. \cite{Yang2015} and discussed in the caption of Fig. \ref{fig2}.} \label{fig1}
\end{figure}

The wavenumber, $k$, describes the spatial distribution of $T(x,t)$ and $I_Q(x,t)$. It has real and imaginary parts that control the decaying and oscillating components of the solutions, respectively, which are plotted versus frequency in Fig. \ref{fig1}(a). At low frequency ($\omega\ll1/\tau_Q$) ${\rm Re}[k]={\rm Im}[k]=\sqrt{\omega/\lambda v_x^+}$, which can be rewritten as $\sqrt{\pi C_V f/ \kappa}$ the well-known classical expression for penetration depth \cite{Regner2013,Koh2007}. At high frequency ($\omega\gg1/\tau_Q$) ${\rm Re}[k]=1/\lambda$ and ${\rm Im}[k]=\omega/v_x^+$, indicating that phonons on average cannot decay on a length scale shorter than the MFP, and that phonons travel at their ballistic velocity $v_x^+$. 

The temperature profiles for low and high frequency are shown in Fig. \ref{fig1}(b)-(c). With the traditional approach, $T(x)$ looks like Fig. \ref{fig1}(b) at all frequencies, while with the HHE the shape changes at higher frequency (Fig. \ref{fig1}(c)). At low frequency transport is diffusive, and at higher frequencies transport becomes quasi-ballistic (purely ballistic when $\lambda\rightarrow \infty$). We find that the key quantity controlling the transition from diffusive to quasi-ballistic transport is the time $\tau_Q$ relative to $1/\omega$, as previously highlighted by Yang and Dames \cite{Yang2015}. It is important to keep in mind that $\tau_Q=\lambda/(2v_x^+)$ varies with dimensionality. In the case of isotropic dispersion and scattering time, we have $\lambda=(4/3)v_g \tau$ in 3D, $(\pi/2)v_g \tau$ in 2D and $2v_g \tau$ in 1D, as well as $v_x^+=v_g/2$ in 3D, $(2/\pi)v_g$ in 2D and $v_g$ in 1D, where $v_g$ is the group velocity and $\tau$ is the phonon scattering time. This gives $\tau_Q= (4/3)\tau$ in 3D, $(\pi^2/8)\tau$ in 2D and $\tau$ in 1D. See Appendix \ref{app:notetauQ} for an alternative formulation of $\tau_Q$.

\begin{figure}	
\includegraphics[width=8.5cm]{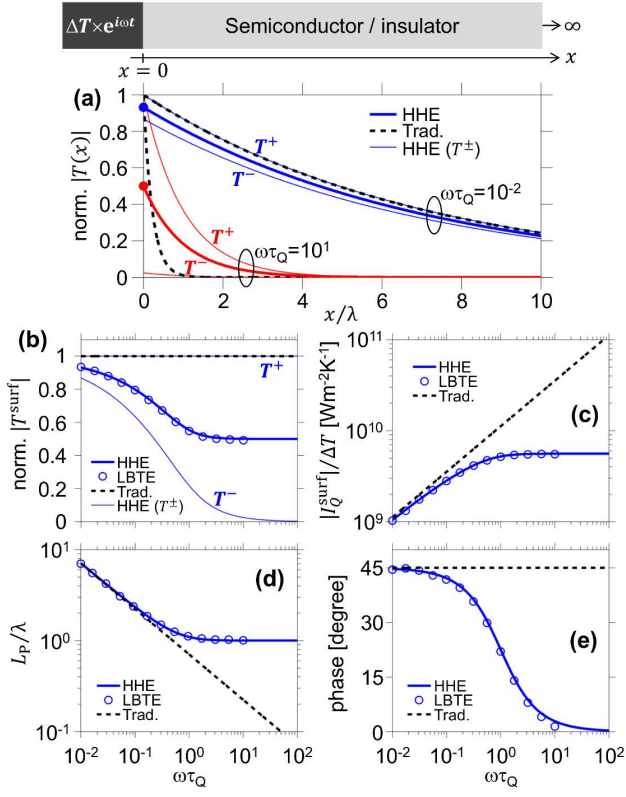}
\caption{Thermal response with the temperature-controlled condition at the surface. (a) Normalized temperature profile $(T(x)-T_0)/\Delta T$ versus normalized position $x/\lambda$ for $\omega \tau_Q$=$10^{-2}$, $10^1$, where $T_0$ is the background temperature. Thick solid lines are solutions to the HHE with the physically correct boundary conditions given by Eq. (\ref{bc:temp}). $T^{\pm}$ indicate the temperature of the forward/backward phonon distributions, with $T=(T^+ + T^-)/2$. The traditional approach corresponds to solving the heat equation. The surface temperature $T^{\rm surf}$ (b), surface heat current $I_Q^{\rm surf}$ (c), penetration depth $L_P$ (d) and phase diffenrence between $T^{\rm surf}$ and $I_Q^{\rm surf}$ are plotted versus frequency $\omega \tau_Q$. Markers are results of the phonon lattice BTE (LBTE, taken from \cite{Yang2015}). We adopted the parameters in \cite{Yang2015} for this 1D problem (see Appendix \ref{app:notetauQ}): $\lambda$=2$\times$41 nm, $v_x^+$=6733 m/s and $C_V$=1.66$\times$$10^6$ Jm$^{-3}$K$^{-1}$.} \label{fig2}
\end{figure}

\subsection{Comparison to the phonon Boltzmann transport equation}
\label{sec:bte}

\subsubsection{Case of temperature-controlled surface}
Using Eq. (\ref{sol:temp_bc1}), Fig. \ref{fig2}(a) shows the normalized temperature profile $(T(x)-T_0)/\Delta T$ ($T_0$ is the background temperature) versus normalized position ($x/\lambda$) for $\omega \tau_Q$=$10^{-2}$, $10^1$. Solutions of our approach, using the HHE with the correct physical boundary conditions (Eq. (\ref{bc:temp})), are shown as thick solid lines. Solutions to the traditional approach, using the heat equation (HE) with the traditional boundary condition $T^{\rm HE}(0^+,t)=\Delta T e^{i\omega t}$, are shown as dashed lines. 

At the lower frequency both our approach and the traditional approach yield similar temperature profiles, while at higher frequency we observe significant deviations near the surface. At the higher $\omega \tau_Q$, and shorter time scales, phonons do not have sufficient time to scatter enough to achieve near local equilibrium. Thus, the phonons travel quasi-ballistically which leads to a nonequilibrium phonon distribution.

To visualize the out-of-equilibrium phonon population, we plot the temperature profiles of the forward and backward moving phonons $T^{\pm}(x)=T(x) \pm R_{\rm th}^{\rm ball}I_Q(x)/2$, where $K^{\rm ball}=1/R_{\rm th}^{\rm ball}=C_V v_x^+/2$ is the ballistic thermal conductance \cite{Maassen2015a,Maassen2015b}. It is important to note that our derivation of the HHE and Cattaneo eq. starting from the McKelvey-Shockley equations does not assume local thermal equilibrium \cite{Maassen2015b}. Allowing both halves of the phonon population to be different and to have separate temperatures $T^{\pm}$ is key to capturing ballistic transport effects. The temperature that appears in Eq. (\ref{hhe}) is simply the average of both forward and reverse temperatures, $T=(T^+ + T^-)/2$ \cite{Maassen2015a,Maassen2015b}. When $T^+$ and $T^-$ are close, the phonons are near equilibrium and transport is diffusive. When there is a large splitting between $T^+$ and $T^-$, the phonons are out of equilibrium and transport is quasi-ballistic, as seen for $\omega \tau_Q$=10$^1$.

The forward-moving phonons injected at the surface are in equilibrium with the contact, $T^+(0^+,t)=\Delta T e^{i\omega t}$, but $T^-(0^+)$ depends on how many injected phonons have time to scatter and return to the surface as backward-moving phonons. As frequency increases, this probability decreases, along with $T^-(0^+)$. This explains the temperature jump observed at $x$=0$^+$, since temperature is the average of both streams $T=(T^+ + T^-)/2$. This is mathematically equivalent to the case of an interface resistance equal to half the ballistic thermal resistance $R_{\rm th}^{\rm ball}I_Q(0^+)/2$, although we consider ideal reflectionless contacts.

In Fig. \ref{fig2} (b)-(e) we present the surface temperature $T^{\rm surf}$, surface heat current $I_Q^{\rm surf}$, penetration depth $L_P$ and phase difference between $T^{\rm surf}$ and $I_Q^{\rm surf}$, versus frequency $\omega \tau_Q$. Thick solid lines are solutions to our approach and markers are numerical results of the phonon lattice BTE (LBTE) (Ref. \cite{Yang2015}). Excellent agreement is observed. The adopted parameters are those from Ref. \cite{Yang2015}.

Tradionally, as $\omega$ increases the penetration depth decreases which drives a higher heat current (from Fourier's law). We see that at higher frequency $L_P$$\rightarrow$$\lambda$ and $I_Q^{\rm surf}$$\rightarrow$$I_Q^{\rm ball}$, where $I_Q^{\rm ball}$ is the ballistic heat current (the largest possible heat current). The phase changes from 45$^{\circ}$ to 0$^{\circ}$ as $\omega$ increases, since only the forward-moving phonons contribute to the temperature and the heat current as transport becomes more ballistic. Thus $T^{\rm surf}$ and $I_Q^{\rm surf}$ both respond instantaneously to the energy injected from the contact, i.e. the heat is carried away from the surface as efficiently as possible. The traditional approach breaks down at high frequency when ballistic effects become important, however the HHE and the Cattaneo eq. are shown to extend the traditional approach to much higher frequencies. 

We note that our solutions, given by Eqns. (\ref{sol:temp_bc1})-(\ref{sol:iq_bc1}), are identical to those reported by Yang and Dames \cite{Yang2015} for a 1D problem. In fact by adding their flux equations derived from the BTE (Eqns. (A6)-(A7)) we directly obtain the HHE, thus indicating that a solution of the 1D BTE is equivalent to that of the HHE. For a 3D material, where the angle-dependence of phonon transport must be considered, both approaches show that the 1D equations remain valid with a rescaling of the input parameters (as we discussed above), which leads to some small numerical differences. Our results are also identical to those reported by Regner et al. \cite{Regner2014} if we multiply our thermal relaxation time by 3/4, which is equivalent to $\tau=(3/4)\tau_Q$ the phonon scattering time. These differences originate in how the angle-dependence of phonon transport is approximated; we replace the $x$-projected phonon velocity distribution with its angle-averaged value, $v_x^+$, while in Refs. \cite{Regner2014,Yang2015} solutions come from taking the two lowest order moments of the BTE. Note that an isotropic medium assumption is not required, and that the input parameters $v_x^+$ and $\lambda$ can be extracted for any phonon dispersion and scattering time (see Eqns. (\ref{eq:lambda_gen})-(\ref{eq:vplus_gen})).

\begin{figure}	
\includegraphics[width=6.5cm]{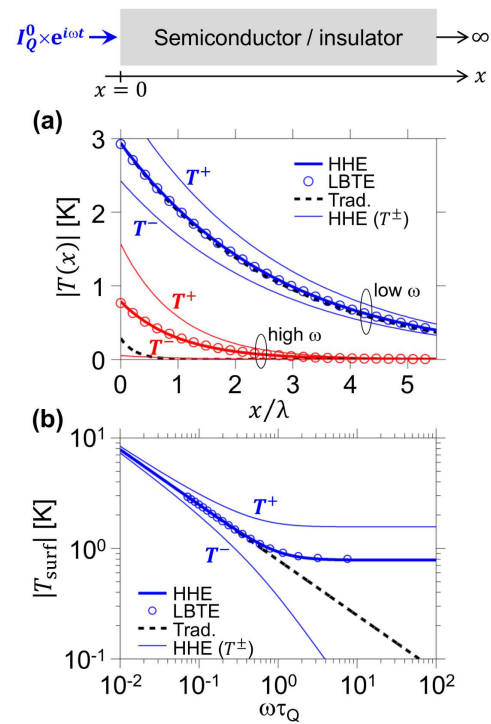}
\caption{Thermal response with the heat current-controlled condition at the surface. (a) Temperature profile $T(x)$ versus normalized position $x/\lambda$ for $f$=$\omega/2\pi$=7.2$\times$10$^{8}$ Hz and 7.6$\times$10$^{10}$ Hz. (b) Surface temperature $T^{\rm surf}$ versus frequency $\omega \tau_Q$. Thick solid lines are solutions to the HHE with the physically correct boundary conditions given by Eq. (\ref{bc:iq}). Markers are results of the phonon LBTE (taken from \cite{Regner2013}). $T^{\pm}$ indicate the temperature of the forward/backward phonon distributions, with $T=(T^+ + T^-)/2$. The traditional approach corresponds to solving the heat equation. We adopted the parameters in \cite{Regner2013} for this 1D problem using effective 3D parameters (see Appendix \ref{app:notetauQ}): $\lambda$=(4/3)$\alpha$$\times$40 nm, $v_x^+$=6733/(2$\alpha$) m/s and $C_V$=1.66$\times$$10^6$ Jm$^{-3}$K$^{-1}$. We find that $\alpha$=$\sqrt{2}$ reproduces the numerical results of the BTE in \cite{Regner2013}. This is equivalent to multiplying $\tau_Q$ in Eq. (\ref{hhe}) by two, which is likely due to a difference in definition of $\tau_Q$.} \label{fig3}
\end{figure}

\subsubsection{Case of heat current-controlled surface}
Using Eq. (\ref{sol:temp_bc2}), Fig. \ref{fig3}(a) shows the temperature profile $T(x)$ versus normalized position ($x/\lambda$) for two frequencies. Solutions of our approach are shown as thick solid lines. Markers are numerical results of the LBTE (Ref. \cite{Regner2013}). Excellent agreement is observed. The adopted parameters are those from Ref. \cite{Regner2013}, but we found that we had to multiply $\tau_Q$ in Eq. (\ref{hhe}) by two, which is probably due to a different definition of $\tau_Q$. A significant splitting in $T^+$ and $T^-$ is observed, a signature of nonequilibrium phonons arising from ballistic effects, which becomes more pronounced at higher frequency.

Fig. \ref{fig3}(b) presents the surface temperature $T^{\rm surf}$ versus $\omega \tau_Q$. Contrary to the temperature-controlled case, the heat current-controlled case gives a temperature that is larger than that expected from the traditional approach. The heat current can be written as $I_Q = K^{\rm ball} (T^+ - T^-)$, thus the magnitude of $T^+ - T^-$ at the surface is constant. At higher frequencies the contribution to $T^-$ decreases, since injected phonons are less likely to scatter on short time scales and return to the surface, and only the excess forward-moving phonons carry the heat current. This requires that $I_Q(0^+)$, the magnitude of which is a constant, approaches $K^{\rm ball} \delta T^+(0^+)$ as $\omega\gg 1/\tau_Q$, where $\delta T^+$ is the excess temperature variation around the background temperature.

\subsubsection{Apparent thermal conductivity}
It is common to define an apparent thermal conductivity, $\kappa_{\rm app}$, that captures the effect of non-diffusive phonon transport through a reduction in bulk thermal conductivity as ballistic effects become prominent. $\kappa_{\rm app}$ is defined as the thermal conductivity that is extracted by assuming heat transport can be described by traditional Fourier's law and heat equation. Using the definition $\kappa_{\rm app} = |I_Q/(-\partial T/\partial x)|$ \cite{Yang2015} with our solutions for $T(x,t)$ and $I_Q(x,t)$ we obtain 
\begin{align}
\kappa_{\rm app} = \kappa_{\rm bulk}/\sqrt{1+(\omega\tau_Q)^2}, \label{eq:app_kappa} 
\end{align}
where the $x$ and $t$ dependences cancel out. This expression is insensitive to the choice of boundary type (i.e. temperature-controlled or heat current-controlled). We find $\kappa_{\rm app}\rightarrow\kappa_{\rm bulk}$ when $\omega\tau_Q<<1$, as expected, and $\kappa_{\rm app}\rightarrow \kappa_{\rm bulk}/(\omega\tau_Q)$ when $\omega\tau_Q>>1$. This simple equation for $\kappa_{\rm app}$ is identical to that reported by Yang and Dames \cite{Yang2015}, although both expressions appear different. In Ref. \cite{Yang2015} a frequency of interest is defined as when $\kappa_{\rm app}=\kappa_{\rm bulk}/2$, which is numerically determined to be $\omega\tau_Q=1.73$. A straightforward evaluation of Eq. (\ref{eq:app_kappa}) shows this condition corresponds to $\omega\tau_Q=\sqrt{3}\approx 1.73$. 

Using the same definition for $\kappa_{\rm app}$, our results are consistent with those reported by Regner et al. \cite{Regner2014}, given that our solutions for $T(x,t)$ and $I_Q(x,t)$ in 3D are identical if we replace $\tau_Q\rightarrow\tau$ (as discussed above). The suppression function in this case is given by $S = \kappa_{\rm app}/\kappa_{\rm bulk} = 1/\sqrt{1+(\omega\tau_Q)^2}$. This result suggests that it may be more convenient, in the case of this particular model problem, to integrate over heat relaxation time to obtain the apparent thermal conductivity as opposed to the mean-free-path, i.e. $\kappa_{\rm app}(\omega)=\int_0^{\infty} S(\omega,\tau_Q) \, \kappa_{\rm bulk}(\tau_Q) \, {\rm d}\tau_Q$ (a point highlighted by Yang and Dames \cite{Yang2015}). In general, mean-free-path may be the more convenient integration quantity, for example when including the effect of laser spot size.

\subsection{Full phonon dispersion and mean-free-path distribution: case of bulk silicon}
\label{sec:full}
The results presented up to this point have been within the gray approximation, considering only a single phonon velocity and MFP. In realistic materials, the full phonon dispersion and energy-dependent MFP distribution must be treated. The most straight-forward extension is to consider the phonon energy (frequency) channels as independent. When deriving the McKelvey-Shockley flux equations, with scattering treated at the level of the relaxation time approximation, the energy channels decouple. This is clearly an approximation and concerns have been raised \cite{Bjorn2016}, but comparisons to full solutions of the phonon BTE in the steady-state \cite{Maassen2015a} and transient \cite{Maassen2015b} cases show reasonable agreement. In this section we demonstrate how our approach can treat realistic materials by using analytical solutions at each energy and then performing the appropriate integration over all energy channels.

As a case study, we consider bulk silicon with the full phonon dispersion extracted from first principles calculations, and including boundary, defect and phonon-phonon Umklapp scatterings treated with phenomenological models calibrated to experimental data. This model provides good agreement with both the experimental phonon energies and the measured temperature dependence of the thermal conductivity. Details can be found in Refs. \cite{Maassen2015a,Maassen2015b}. Using the detailed material properties of Si we extract $v_x^+(\epsilon)$ and $\lambda(\epsilon)$ (see Appendix \ref{app:notetauQ}), which are used to evaluate $k(\epsilon)$, $\kappa(\epsilon)$ and $\tau_Q(\epsilon)$ appearing in our solutions of $T$ (Eq. (\ref{sol:temp_bc1}) and Eq. (\ref{sol:temp_bc2})) and $I_Q$ (Eq. (\ref{sol:iq_bc1}) and Eq. (\ref{sol:iq_bc2})).

\begin{figure}	
\includegraphics[width=6.5cm]{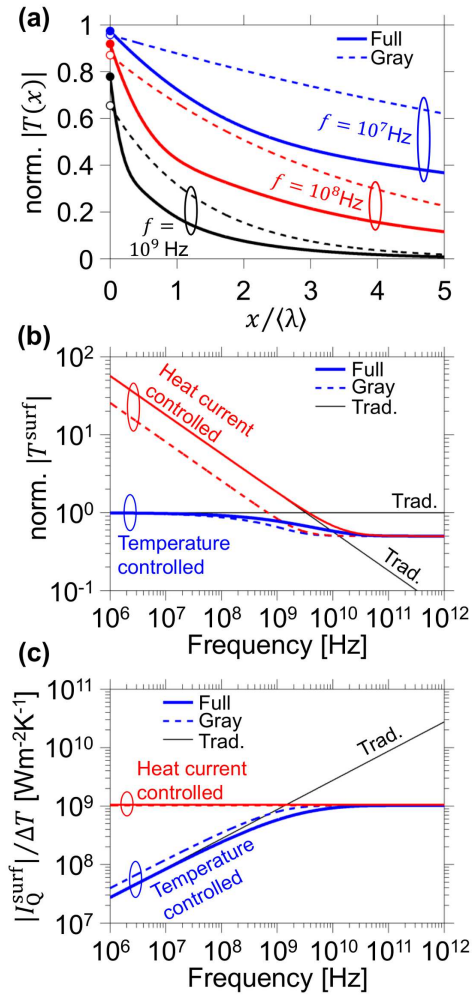}
\caption{Thermal response of bulk silicon using full phonon dispersion and mean-free-path distribution. (a) Normalized temperature profile $(T(x)-T_0)/\Delta T$ versus normalized position $x/\lambda$ for $f$=$\omega/2\pi$=10$^7$Hz, 10$^8$Hz and 10$^9$Hz, where $T_0$ is the background temperature. A temperature-controlled condition is used. (b)-(c) Surface temperature $T^{\rm surf}$ and surface heat current $I_Q^{\rm surf}$ versus frequency $\omega \tau_Q$. Thick solid lines are solutions to the HHE and Cattaneo equation. Dashed lines correspond to the gray approximation using $\lambda$=151 nm, $v_x^+$=1255 m/s and $C_V$=1.63$\times$10$^6$ Jm$^{-3}$K$^{-1}$. Thin solid lines are solutions to the heat equation and Fourier's law. We used the same full dispersion and mean-free-path distribution of bulk silicon as Refs. \cite{Maassen2015a,Maassen2015b}.} \label{fig4}
\end{figure}

We compute $T(x,t,\epsilon)$ and $I_Q(x,t,\epsilon)$ at each energy, and obtain the total temperature and heat current using \cite{Maassen2015a,Maassen2015b}:
\begin{align}
 T(x,t) &= \int_0^{\infty} T(x,t,\epsilon)\, C_V(\epsilon) \,{\rm d}\epsilon \,/ C_V, \label{eq:temp_en}\\
 I_Q(x,t) &= \int_0^{\infty} I_Q(x,t,\epsilon) \, {\rm d}\epsilon, \label{eq:iq_en}
\end{align}
where $C_V(\epsilon)=\epsilon\, D(\epsilon)\, [\,\partial n_0/\partial T\,]$ is the energy-dependent heat capacity, $D(\epsilon)$ is the phonon density of states, $n_0$ is the equilibrium Bose-Einstein distribution evaluated at the reference temperature $T_0$, and $C_V = \int_0^{\infty} C_V(\epsilon)\, {\rm d}\epsilon$ is the total heat capacity. Since each phonon energy is assumed to be independent, our analytical solutions can be evaluated at each energy with the final results obtained from the appropriate energy integration provided by Eqns. (\ref{eq:temp_en})-(\ref{eq:iq_en}).

Fig. \ref{fig4}(a) shows the temperature profile $T(x)$ versus normalized position ($x/\lambda$) in bulk Si for $f=\omega/2\pi$=10$^7$, 10$^8$, 10$^9$ Hz. Full solutions to our approach (solid lines) are compared to solutions evaluated at a single energy using the average phonon properties, i.e. the gray approximation (dashed lines). Small differences between the full and gray results are observed at the surface; larger differences occur inside the material. This indicates that the gray approximation within this approach can lead to large errors. Another observation, most clearly seen at higher frequency, is that the full solutions do not yield exponential temperature profiles.

Fig. \ref{fig4}(b)-(c) presents the surface temperature $T^{\rm surf}$ and heat current $I_Q^{\rm surf}$ versus frequency, showing both cases of temperature- and heat current-controlled conditions. We compare full solutions (thick solid lines) to the gray approximation (dashed lines) and the traditional approach (thin solid lines). When considering surface properties, the full and gray solutions are reasonably close and within roughly a factor of two. The surface temperature and heat current are sensitive to the choice of boundary type ($T$- versus $I_Q$-controlled), and the differences increase as the frequency decreases. This is most prominent at lower frequencies where transport is diffusive, and is, in fact, well-known from classical thermal physics. We also note that although the solutions appear different the thermal conductivity and thermal resistance are invariant to the choice of either $T$- and $I_Q$-controlled cases. Later we highlight how ballistic effects do lead to differences in the solutions for both cases.

\section{Discussion}
\label{sec:discussion}
Fig. \ref{fig4}(b)-(c) shows that deviations from the traditional approach appear at different frequencies depending on the adopted heating case. We define the diffusive-to-ballistic transition frequency as the frequency at which the full solution differs from the traditional solution by 5\%. In Fig. \ref{fig4}(b) this transition occurs at 3.6$\times$10$^7$ Hz and 3.2$\times$10$^9$ Hz for the $T$- and $I_Q$-controlled cases, respectively, a factor of roughly 10$^2$. Ballistic effects can come in through either the governing equations (i.e. Eqns. (\ref{hhe})-(\ref{cattaneo})) and/or the boundary conditions (i.e. Eqns. (\ref{bc:temp})-(\ref{bc:iq})). The former cannot explain the difference in transition frequency, since we solve the HHE and Cattaneo eq. in both cases. For the $I_Q$-controlled case the boundary condition (Eq. (\ref{bc:iq})) is the same in the traditional limit (i.e. with no ballistic effects), however for the $T$-controlled case the boundary condition (Eq. (\ref{bc:temp})) differs from the traditional limit due to the term with $\partial T/\partial x$. This extra term, responsible for the temperature jump at the surface, drives the full solution away from the traditional solution at a lower frequency compared to the $I_Q$-controlled case. In Fig. \ref{fig4}(c) the transition frequency is 1.4$\times$10$^7$ Hz for the $T$-controlled case, while no deviation from the traditional solution is observed for the $I_Q$-controlled case (boundary condition for both the full and traditional solutions are the same).

When analyzing experimental data, several factors should be considered. For example, it is not clear, which case ($T$- or $I_Q$-controlled) should be used at the surface of the semiconductor. We note, though, that phase lag is a key quantity in the analysis of FDTR and TDTR experiments, and that our results indicate phase is insensitive to the adopted heating type. One-dimensional solutions are also probably not adequate, and it is not clear if and how the metal itself and the metal-semiconductor junction should be treated \cite{Regner2015}. The diffusive-to-ballistic transition frequencies reported here are quite high, which could be a result of the simplified model we used to demontrate the technique.

\section{Summary}
\label{sec:summary}
We have shown that when the correct physical boundary conditions are used, the HHE and Cattaneo equation (i.e. diffusion equations) can be used to model ballistic effects in frequency-dependent transient thermal transport. Our analytical solutions, derived for the case of temperature- and heat current-controlled heating at the surface, are found to reproduce rigorous solutions of the phonon BTE with high accuracy. Numerical solutions of the BTE were available for only 1D transport problems within the gray approximation, so our approach remains to be tested in cases where a treatment of angle-dependent phonon transport is required and scattering is handled beyond the relaxation time approximation.

By calculating the thermal transport response of bulk silicon, we demonstrated how the approach can easily handle a full phonon dispersion and energy-dependent MFP distribution when the energy channels are treated as independent. Testing the gray approximation, we found that it performs resonably well at the semiconductor/insulator surface, but it fails to accurately describe the temperature profile inside the material. 

An advantage of this simple approach based on solving diffusion equations is its physical transparency. For example, we discussed how results can be explained in terms of the directed temperatures, $T^{\pm}$, (the temperature of each half of the phonon distribution) and how a large splitting of $T^+$ and $T^-$ results from the nonequilibrium nature of ballistic transport. The main conclusion of this work is that diffusion equations have the potential for accurately treating ballistic to diffusive thermal transport.

\acknowledgements
This work was supported in part by DARPA MESO (Grant N66001-11-1-4107) and through the NCN-NEEDS program, which is funded by the National Science Foundation, contract 1227020-EEC, and the Semiconductor Research Corporation.

\appendix
\section{Boundary conditions at the surface}
\label{app:BC}
Here we derive the correct physical boundary conditions, that come from the phonon BTE, to be used when solving the HHE and the Cattaneo equation. Below we consider two cases: when heating at the surface is driven by a controlled temperature or a controlled heat current. Our starting point are the directed heat fluxes $I_Q^{\pm}$ used in the McKelvey-Shockley equations, and how to rewrite the boundary conditions for $I_Q^{\pm}$ into boundary conditions for $T$ and $I_Q$. Instead of dealing with $I_Q^{\pm}$ we can work with the directed phonon temperatures $T^{\pm}=\delta I_Q^{\pm}/K^{\rm ball}+T_0$ \cite{Maassen2015a,Maassen2015b} (i.e. temperature of each half of the phonon distribution), where $\delta I_Q^{\pm}$ is the variation in directed heat flux around the background equilibrium flux and $T_0$ is the background temperature. 

We begin by writing two general expressions that relate $T^{\pm}$ to $T$ and $I_Q$ \cite{Maassen2015a,Maassen2015b}: 
\begin{align}
 T(x,t) &= \left[ T^+(x,t) + T^-(x,t) \right] / 2, \\
I_Q(x,t) &= K^{\rm ball} \left[ T^+(x,t) - T^-(x,t) \right].
\end{align}
By eliminating $T^-$, we find
\begin{align}
 T^+(0^+,t) &= T(0^+,t) + \frac{I_Q R_{\rm th}^{\rm ball}}{2}. 
\end{align}
Using $K^{\rm ball} = 1/R_{\rm th}^{\rm ball} = C_V v_x^+ /2$ \cite{Maassen2015a,Maassen2015b} with the Cattaneo equation, we obtain
\begin{align}
 T(0^+,t) - \frac{\lambda}{2\left( 1+i\omega \tau_Q \right)}\left.\frac{\partial T}{\partial x}\right|_{0^+} &= T^+(0^+,t). 
\end{align}
For the temperature-controlled case, where the surface of the semiconductor/insulator is joined to a thermal reservoir via an ideal reflectionless contact, the temperature of the injected phonons is equal to the temperature of the reservoir. This gives the boundary condition that is Eq. (\ref{bc:temp}). For the heat current-controlled case, the HHE and Cattaneo equation can be solved using $I_Q(0^+,t)=I_Q^0 e^{i\omega t}$ (Eq. (\ref{bc:iq})), which states that $|T^+(0^+,t)-T^-(0^+,t)|$ must be a constant.

\section{Definition of $\tau_Q$}
\label{app:notetauQ}
As shown in the main text and in Ref. \cite{Maassen2015b} the thermal relaxation time is written as $\tau_Q = \lambda/(2v_x^+)$. The mean-free-path for backscattering and average $x$-projected velocity are defined as \cite{Maassen2015a,Maassen2015b}
\begin{align}
\lambda(\epsilon) &= 2 \frac{\langle v_x^2(k)\, \tau(k) \rangle}{\langle |v_x(k)| \rangle}, \label{eq:lambda_gen}\\ 
v_x^+(\epsilon) &= \langle |v_x(k)| \rangle, \label{eq:vplus_gen}
\end{align} 
where $\tau$ is the phonon scattering time, $\langle X \rangle = \sum_k X(k) \delta(\epsilon-\epsilon(k))/\sum_k \delta(\epsilon-\epsilon(k))$, $\epsilon(k)$ is the phonon energy dispersion relation (related to the phonon frequency dispersion through $\nu(k)=\epsilon(k)/h$), $k$ is a vector in reciprocal space, and the summation over $k$ is restricted to the Brillouin zone. Using the above relations for $\lambda$ and $v_x^+$ and inserting them into our expression for $\tau_Q$ we find
\begin{align}
\tau_Q(\epsilon) &= \frac{\langle v_x^2(k)\, \tau(k) \rangle}{\langle |v_x(k)| \rangle^2}. \label{def:tauQ}
\end{align} 
Thus $\tau_Q$ is related to $\tau$ through Eq. (\ref{def:tauQ}) which depends on the phonon dispersion.



\begin{thebibliography}{00}
%
%
\bibitem{Koh2007} Y. K. Koh and D. G. Cahill, Phys. Rev. B {\bf 76}, 075207 (2007).
%
%
\bibitem{Siemens2010} M. E. Siemens, Q. Li, R. Yang, K. A. Nelson, E. H. Anderson, M. M. Murnane and H. C. Kapteyn, Nat. Mater. {\bf 9}, 26 (2010).
%
%
\bibitem{Minnich2011a} A. J. Minnich, J. A. Johnson, A. J. Schmidt, K. Esfarjani, M. S. Dresselhaus, K. A. Nelson and G. Chen, Phys. Rev. Lett. {\bf 107}, 095901 (2011).
%
%
\bibitem{Johnson2013} J. A. Johnson, A. A. Maznev, J. Cuffe, J. K. Eliason, A. J. Minnich, T. Kehoe, C. M. S. Torres, G. Chen and K. A. Nelson, Phys. Rev. Lett. {\bf 110}, 025901 (2013).
%
%
\bibitem{Regner2013} K. T. Regner, D. P. Sellan, Z. Su, C. H. Amon, A. J. H. McGaughey and J. A. Malen, Nat. Comm. {\bf 4}, 1640 (2013).
%
%
\bibitem{Wilson2014} R. B. Wilson and D. G. Cahill, Nat. Comm. {\bf 5}, 5075 (2014).
%
%



\bibitem{Minnich2011b} A. J. Minnich, G. Chen, S. Mansoor and B. S. Yilbas, Phys. Rev. B {\bf 84}, 235207 (2011).
%
%
\bibitem{Peraud2011} J.-P. M. Peraud and N. G. Hadjiconstantinou, Phys. Rev. B {\bf 84}, 205331 (2011).
%
%
\bibitem{Peraud2012} J.-P. M. Peraud and N. G. Hadjiconstantinou, Appl. Phys. Lett. {\bf 101}, 153114 (2012).
%
%
\bibitem{Collins2013} K. C. Collins, A. A. Maznev, Z. Tian, K. Esfarjani, K. A. Nelson and G. Chen, J. Appl. Phys. {\bf 114}, 104302 (2013).
%
%
\bibitem{Ding2014} D. Ding, X. Chen and A. J. Minnich, Appl. Phys. Lett. {\bf 104}, 143104 (2014).
%
%
\bibitem{Zeng2014} L. Zeng and G. Chen, J. Appl. Phys. {\bf 116}, 064307 (2014).
%
%


\bibitem{Maznev2011} A. A. Maznev, J. A. Johnson and K. A. Nelson, Phys. Rev. B {\bf 84}, 195206 (2011).
%
%
\bibitem{Wilson2013} R. B. Wilson, J. P. Feser, G. T. Hohensee and D. G. Cahill, Phys. Rev. B {\bf 88}, 144305 (2013).
%
%
\bibitem{Regner2014} K. T. Regner, A. J. H. McGaughey and J. A. Malen, Phys. Rev. B {\bf 90}, 064302 (2014).
%
%
\bibitem{Vermeersch2014a} B. Vermeersch, J. Carrete, N. Mingo and A. Shakouri, Phys. Rev. B {\bf 91}, 085202 (2015).
%
%
\bibitem{Vermeersch2014b} B. Vermeersch, A. M. S. Mohammed, G. Pernot, Y. R. Koh and A. Shakouri, Phys. Rev. B {\bf 91}, 085203 (2015).
%
%
\bibitem{Hua2014a} C. Hua and A. J. Minnich, Phys. Rev. B {\bf 89}, 094302 (2014).
%
%
\bibitem{Hua2014b} C. Hua and A. J. Minnich, Phys. Rev. B {\bf 90}, 214306 (2014).
%
%
\bibitem{Yang2015} F. Yang and C. Dames, Phys. Rev. B {\bf 91}, 165311 (2015).
%
%
\bibitem{Hua2015} C. Hua and A. J. Minnich, J. Appl. Phys. {\bf 117}, 175306 (2015).
%
%


\bibitem{Maassen2015a} J. Maassen and M. Lundstrom, J. Appl. Phys. {\bf 117}, 035104 (2015).
%
%
\bibitem{Maassen2015b} J. Maassen and M. Lundstrom, J. Appl. Phys. {\bf 117}, 135102 (2015).
%
%
\bibitem{Collins2014} K. C. Collins, A. A. Maznev, J. Cuffe, K. A. Nelson and G. Chen, {\bf 85}, 124903 (2014).
%
%
\bibitem{Mckelvey1961} J. P. McKelvey, R. L. Longini and T. P. Brody, Phys. Rev. {\bf 123}, 51 (1961).
%
%
\bibitem{Shockley1962} W. Shockley, Phys. Rev. {\bf 125}, 1570 (1962).
%
%
\bibitem{Cattaneo1958} C. Cattaneo, Compte Rendus {\bf 247}, 431 (1958).
%
%
\bibitem{Bjorn2016} B. Vermeersch (private communication, 2016).
%
%
\bibitem{Regner2015} K. T. Regner, L. C. Wei and J. A. Malen, J. Appl. Phys. {\bf 118}, 235101 (2015).







\end{thebibliography}
\end{document}